%% file: KsenofontovISCRA2023-1.1a.tex
\documentclass[
    a4paper,notitlepage,
    twocolumn,
    nofootinbib,
    amsmath,amssymb
]{revtex4-2}

\usepackage{graphicx}
\graphicspath{{figures/}}

\usepackage[usenames,dvipsnames]{xcolor} % link colors tweak

\usepackage[
    colorlinks = true,
    urlcolor = MidnightBlue,
    linkcolor = BrickRed,
    citecolor = NavyBlue
]{hyperref}

\hypersetup{
    pdfauthor = {A. V. Glushkov, L. T. Ksenofontov, K. G. Lebedev, A. Saburov},
    pdftitle = {A direct comparison of muon measurements at the Yakutsk Array and the Pierre Auger Observatory},
    pdfsubject = {ultra-high energy cosmic rays, extensive air showers},
    pdfkeywords = {extensive air showers, energy spectrum, muon puzzle, the Yakutsk array, the Pierre Auger Observatory}
}

\input{quasi-maik}
\input{abbrev.tex}

\begin{document}

\title{A Direct Comparison of Muon Measurements at the Yakutsk Array and the Pierre Auger Observatory}

\author{\bf A.\,V.\,Glushkov\textsuperscript{1)}}
\email{glushkov@ikfia.ysn.ru}

\author{A.\,V.\,Saburov}
\email{vs.tema@gmail.com}

\author{L.\,T.\,Ksenofontov\textsuperscript{1)}}
\email{ksenofon@ikfia.ysn.ru}

\author{K.\,G.\,Lebedev\textsuperscript{1)}}
\email{LebedevKG@ikfia.ysn.ru}

\affiliation{\normalfont{\textsuperscript{1)}Yu.\,G.\,Shafer Institute of Cosmophysical Research and Aeronomy of Siberian branch of the Russian Academy of Sciences, Yakutsk, Russia}}

\begin{abstract}
    \noindent
    {\bf Abstract~---} Here we consider the results of direct measurements of muons in extensive air showers with zenith angles $\theta \le 45\degr$ and energy above $10^{17}$~eV, obtained at the Pierre Auger Observatory and Yakutsk array. In both experiments muons were registered with underground scintillation detectors with $\approx 1.0 \times \sec\theta$~GeV energy threshold. Measured density values were compared to theoretical predictions calculated within the framework of the \qgsii{} hadron interaction model. They differ by factor $1.53 \pm 0.13$(stat). We demonstrate that this difference is due to overestimation of muon densities by 1.22 times and underestimation of primary energy by 1.25 times in the Auger experiment.
\end{abstract}

\maketitle

\input{01-intro}

\input{02-yakutsk}

\input{03-auger}

\input{04-finale}

\section*{Funding}

This work was made within the framework of the state assignment No.\,122011800084-7 using the data obtained at The Unique Scientific Facility ``The D.\,D.~Krasilnikov Yakutsk Complex EAS Array'' (YEASA) (\url{https://ckp-rf.ru/catalog/usu/73611/}).

\section*{Conflict of interest}

The authors of this work declare that they have no conflict of interest.

\acknowledgements

Authors express their gratitude to the staff of the Separate structural unit YEASA of ShICRA SB RAS.

\bibliographystyle{apsmaik4-2}
\bibliography{KsenofontovISCRA2023}

\end{document}

%% file: quasi-maik.tex
\usepackage{etoolbox}
\makeatletter
\patchcmd{\frontmatter@RRAP@format}{(}{}{}{}
\patchcmd{\frontmatter@RRAP@format}{)}{}{}{}
\makeatother

\newcommand{\eqn}[1]{(\ref{#1})}
\newcommand{\fign}[1]{Fig.~\ref{#1}}
\newcommand{\nthick}{\thickmuskip=0mu} % kill extra space after `\sim` `\simeq` `\ge` `\le`...
\newcommand{\nmed}{\medmuskip=0mu} % tighten up math expressions like (1-5)x10^{17}
\def\thesection       {\arabic{section}}
\def\p@section        {}
\def\thesubsection    {\thesection.\arabic{subsection}}
\def\p@subsection     {\thesection.}

\def\p@subsubsection  {\thesection\,\thesubsection\,}

\makeatletter
\renewcommand \thefigure{\@arabic\c@figure}
\renewcommand \thetable{\@arabic\c@table}
\usepackage{array, multirow}

\newcolumntype{C}[1]{>{\centering\arraybackslash}p{#1}}   % centered
\newcolumntype{R}[1]{>{\raggedleft\arraybackslash}p{#1}}  % align right
\newcolumntype{L}[1]{>{\raggedright\arraybackslash}p{#1}} % align left

%% file: abbrev.tex
\newcommand{\yeas}{Yakutsk}

\newcommand{\E}{E_0}
\newcommand{\fmu}{f_{\mu}}
\newcommand{\Rc}{r_{\text{est.}}}
\newcommand{\ZenC}{\theta_{\text{est.}}}
\newcommand{\RhoC}{\rho_{\text{SD}}(\Rc)}
\newcommand{\Ecut}{\epsilon_{\text{cut}}}

\newcommand{\usec}{$\mu$s}    % usec
\newcommand{\sqrm}{m$^2$}     % m^2
\newcommand{\cubcm}{cm$^3$}   % cm^3
\newcommand{\depth}{g/cm$^2$} % g/cm^2
\newcommand{\dens}{g/cm$^3$}  % g/cm^3
\newcommand{\degr}{^{\circ}}  % degrees

\newcommand{\edep}{\Delta\epsilon}  % total energy deposit in scintillator
\newcommand{\VEM}{\epsilon_1}       % energy deposit of a single vertical muon
\newcommand{\rhoY}{\rho^{\text{Y}}} % Yakutsk particle density
\newcommand{\CCTr}{C$_2$}           % C2 trigger

\newcommand{\rhoPAO}{\rho_{35}}
\newcommand{\rhoPAOT}{\rho_{\theta}}
\newcommand{\fatta}{f_{\text{att.}}^{35}(\theta)}
\newcommand{\fattb}{f_{\text{att.}}^{38}(1000,\theta)}
\newcommand{\Esd}{E_{\text{SD}}}

\newcommand{\rhosSOOT}{\rho_{\text{SD}}(600,\theta)}
\newcommand{\rhosSOOV}{\rho_{\text{SD}}(600,0\degr)}

\newcommand{\xobs}{x_{\text{obs}}}

\newcommand{\qgsii}{{\sc qgsj}et-{\sc ii}.04}

\newcommand{\eposlhc}{{\sc epos-lhc}}

\newcommand{\corsika}{{\sc corsika}}

%% file: 01-intro.tex
\section{Introduction} \label{sec:1}

\begin{figure}[!htb]
    \centering
    \includegraphics[width=0.49\textwidth]{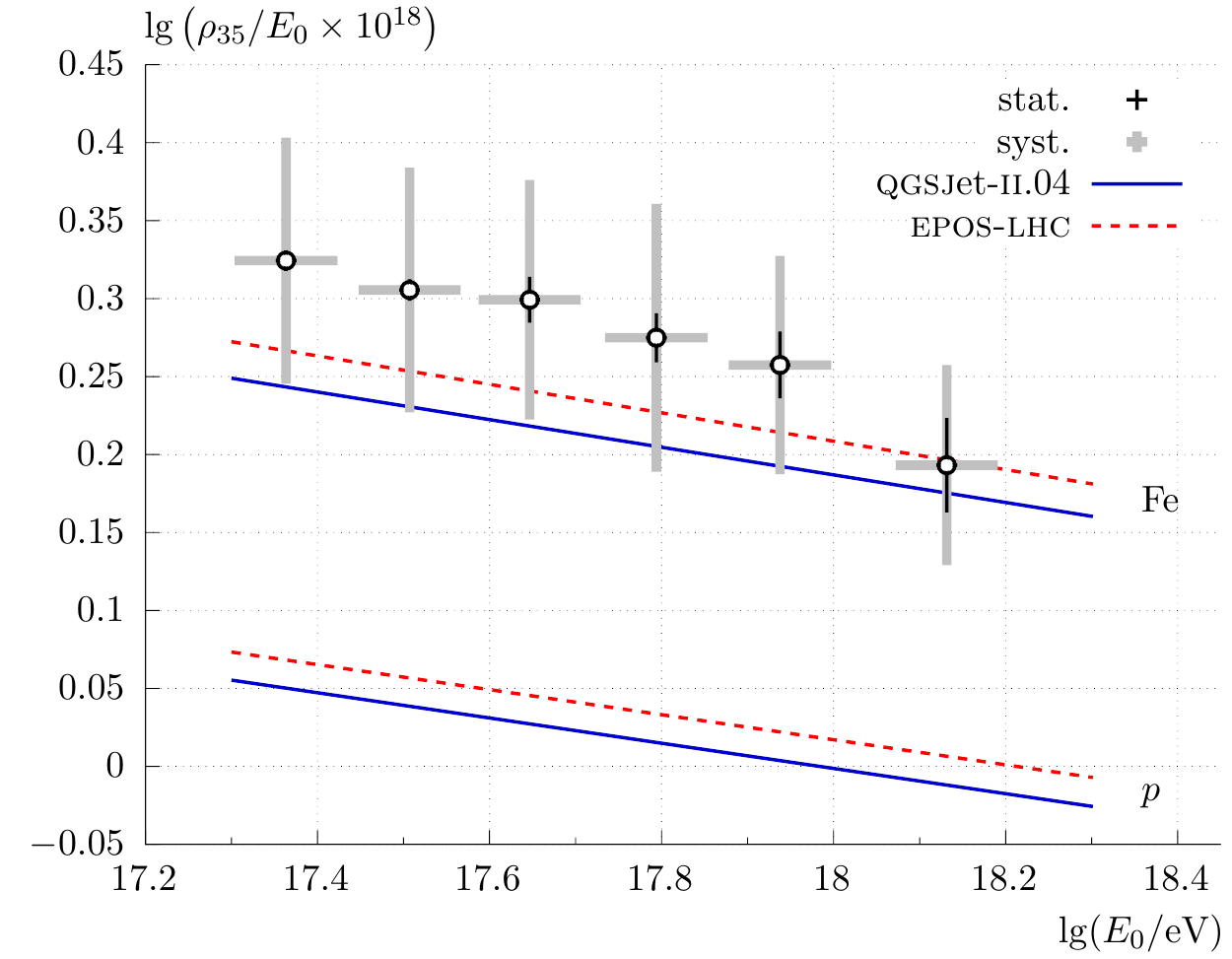}
    \caption{Muon densities in EAS at 450~m from the axis normalized by primary energy $\E$. Results of direct measurements performed at Auger with underground scintillation detectors with $\approx 1.0 \times \sec 35\degr$~GeV threshold. The data were extracted from Fig.~11 in work~\cite{bib:1}.}
    \label{fig:1}
\end{figure}

Recently, the Pierre Auger collaboration (Auger) has reported on direct measurements of muon density in extensive air showers (EAS) from cosmic rays (CR) with energies $2 \times (10^{17} - 10^{18})$~eV and zenith angles $\theta \le 45\degr$~\cite{bib:1}. The measurements were made with 5-\sqrm{} and 10-\sqrm{} scintillation detectors placed under a 2.3-m layer of ground forming a $\approx 1.0~\text{GeV} \times \sec\theta$ registration threshold. The main analyzed parameter was $\rhoPAO$~--- muon density measured in individual events at 450~m from shower axis that was subsequently converted to zenith angle $35\degr$ with the following relations:

\begin{gather}
    \rhoPAO = \rhoPAOT / \fatta\text{,} \\
    \label{eq:1}
    \fatta = 1 + (0.54 \pm 0.10) \cdot x + \nonumber \\
        \qquad + (1.02 \pm 0.69) \cdot x^2\text{,}
    \label{eq:2}
\end{gather}

\noindent
where $x = \cos^2\theta - \cos^2 35\degr$. One of the results from work~\cite{bib:1} is presented in \fign{fig:1}. The obtained values point at abnormally high muon content in EAS when compared to model predictions and do not exclude that the analyzed events have originated from primary iron nuclei. This feature of muon component was considered by international group in a combined analysis based on the data from eight EAS arrays: EAS-MSU, IceCube, KASCADE-Grande, NEVOD-DECOR, Auger, SUGAR, Telescope Array (TA) and Yakutsk Complex EAS Array (\yeas)~\cite{bib:2, WHISP:2021}. It is seen from \fign{fig:1} that the Auger data do not agree with results of \yeas~\cite{bib:2, WHISP:2021, bib:3, bib:4, bib:5, bib:6}. It is hard not to notice that muon component in the Auger experiment~\cite{bib:1} was measured with a similar method that is used at \yeas. This provides the possibility to make a direct comparison between muon measurement techniques in both experiments in order to find sources of existing disagreements.

%% file: 02-yakutsk.tex
\section{Measurement of EAS particle density at Yakutsk array} \label{sec:2}

\subsection{Scintillation Detector} \label{sec:2.1}

\begin{figure}[!htb]
    \centering
    \includegraphics[width=0.49\textwidth]{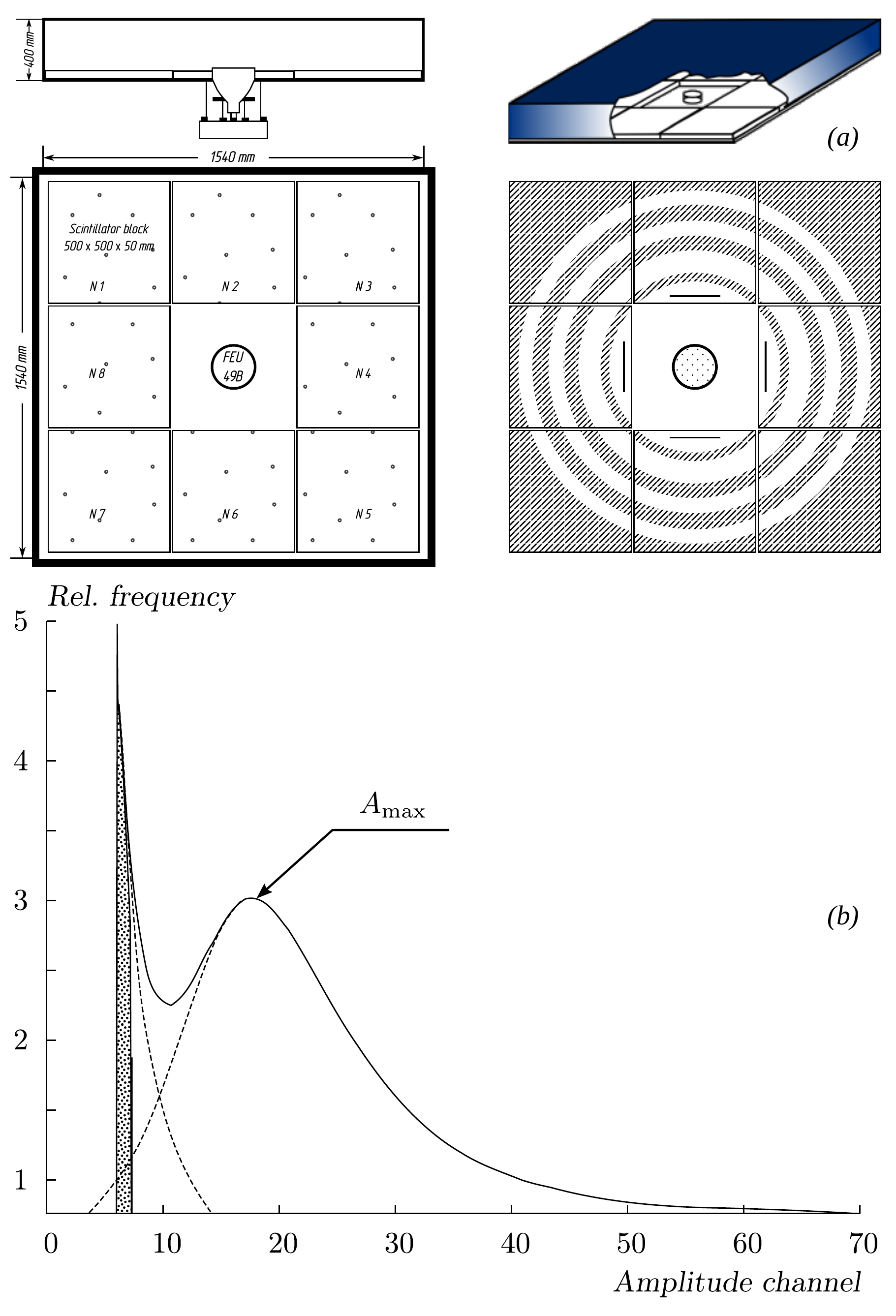}
    \caption{({\sl a}) Standard scintillation detector used at Yakutsk array with $0.25 \times 8 = 2$~\sqrm{} area. A PMT FEU-49 is mounted in the center. ({\sl b}) Differential response spectrum of 2-\sqrm{} scintillation detector from background cosmic muons.}
    \label{fig:2}
\end{figure}

For measurement of particle flux in EASs the \yeas{} experiment utilizes 2-\sqrm{} detectors based on plastic scintillators made of polystyrene (density $1.06$~\dens) with luminescent additives ($\nthick\sim 2$\% p-terphenyl and $\nthick\sim 0.02$\% POPO) in the form of $50 \times 50 \times 5$-\cubcm{} blocks. Eight such blocks are placed around the perimeter of the platform of a light-proof container. In the center of the platform a photomultiplier tube (PMT) FEU-49 is mounted. The cover of the container is made of 1.5-mm aluminium sheet, its inner surfaces are covered with a special white paint. For a more uniform light collection, a pattern of concentric rings is painted on the upper surface of scintillator blocks~--- it absorbs the light in the central area (see \fign{fig:2}{\sl a}). In the future the application of the pattern was abandoned due to its insignificant effect on the measured particle densities.

The light generated inside scintillator blocks from passing particles diffusively reflects from inner surfaces of the cover and is collected by PMT. The glow duration of scintillator is about several nanoseconds, the maximum light yield is approximately at $440$~nm which fits well within spectral characteristics of FEU-49 and reflective properties of painted surfaces of the container.

\subsection{Calibration of Detectors} \label{sec:2.2}

Routine calibration and operation control of detectors are performed using background flux of secondary cosmic particles, which at sea level predominately consists of muons. Individual background muons have rather high energy and a well-determined zenith-angular distribution. At sea level a 2-\sqrm{} detector registers about 400 events per second. A typical response spectrum of uncontrolled detector from background cosmic particles is shown on \fign{fig:2}{\sl b}. Steeply falling branch of the spectrum at low amplitudes arises from noise produced in the PMT itself, weak flashes from low-energy atmospheric particles and radioactive impurities if they are presented in materials of the detector. The position of the $A_{\text{max}}$ value on \fign{fig:2}{\sl b} is close to the $U_1$ amplitude (the response unit) on the PMT anode from passing of a single vertical muon. When $n$ particles traverse through a detector, the amplitude on the PMT anode is equal to $U = n \cdot U_1$.

\subsection{Measurement Units} \label{sec:2.3}

\begin{figure}[htb]
    \centering
    \includegraphics[width=0.49\textwidth]{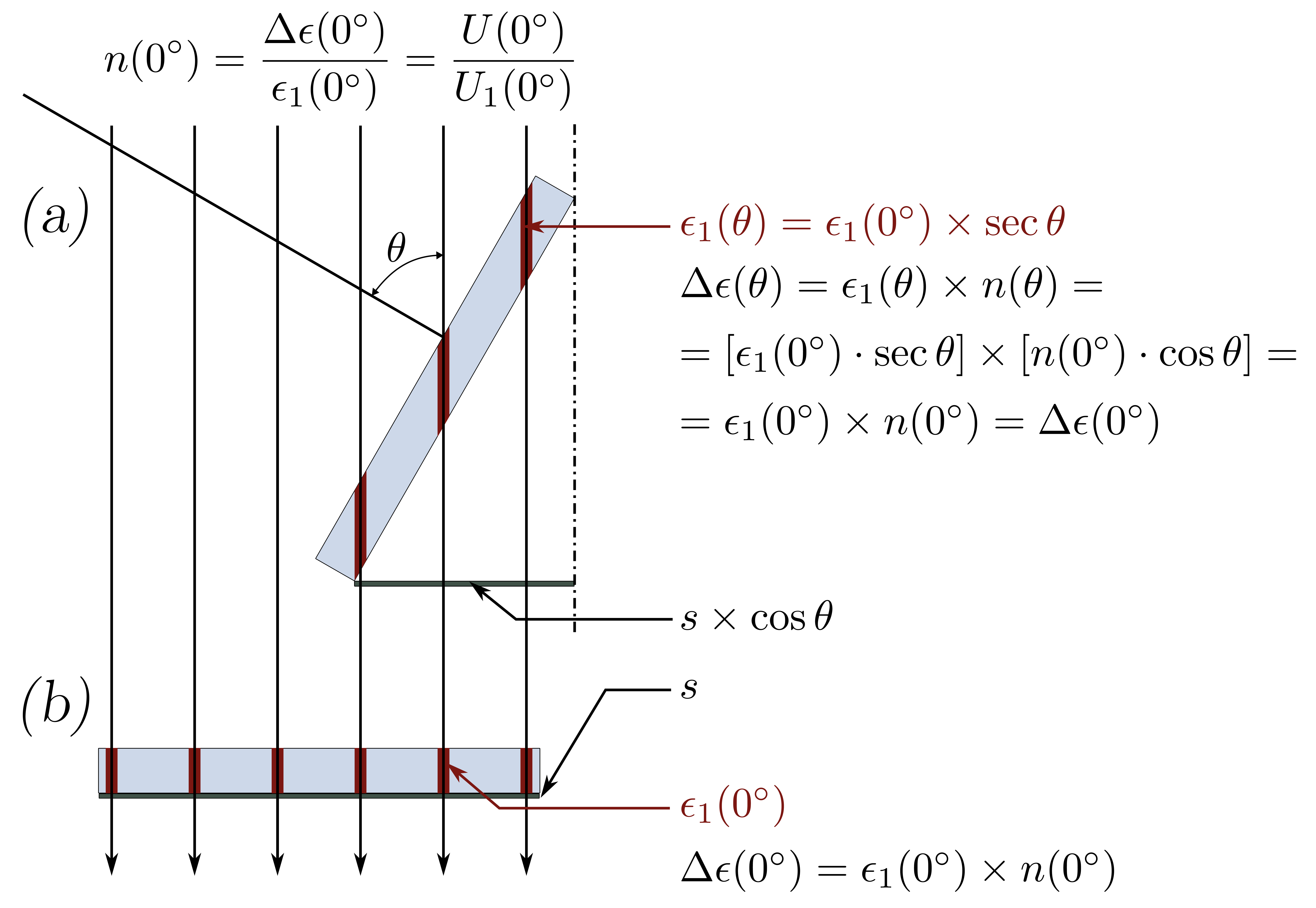}
    \caption{A diagram of total response $\edep$ formation during the passage of a given number of particles $n$ through a scintillation detector of area $s$ in EASs with different zenith angles $\theta$.}
    \label{fig:3}
\end{figure}

Densities of surface and muon EAS components at Yakutsk array are measured in units of energy deposited in 5-cm thick plastic scintillator by vertical relativistic muons. Inside plastic a relativistic particle loses approximately $2.217$~MeV per 1~\depth{} unit of path. The total value $\VEM(0^{\circ}) = 5~\text{cm} \times 1.06~\text{\dens} \times 2.217~\text{MeV~cm}^2/\text{g} = 11.75$~MeV (shown in \fign{fig:3} inside a scintillator block with darkened tracks) is spent on ionization and is re-emitted by scintillator as a flash of light. This flash is subsequently converted by PMT into electric pulse with amplitude $U_1(0\degr)$~--- the single particle level (calibration level). The calibration level is monitored via amplitude spectra which are continuously accumulated from a scintillation detector. Summary energy deposits in inclined ($\edep(\theta)$) and vertical ($\edep(0\degr)$) showers are shown in \fign{fig:3}{\sl a} and \fign{fig:3}{\sl b} accordingly. It is seen that these energy deposits are the same at any zenith angle. The number of particles at axis distance $r$ is defined as

\begin{equation}
    n(r,0\degr) =
    \frac{\edep(r,0\degr)}{\VEM(r,0\degr)} = 
    \frac{U(r,0\degr)}{U_1(r,0\degr)}\text{.}
    \label{eq:3}
\end{equation}

\begin{figure}[!htb]
    \centering
    \includegraphics[width=0.49\textwidth]{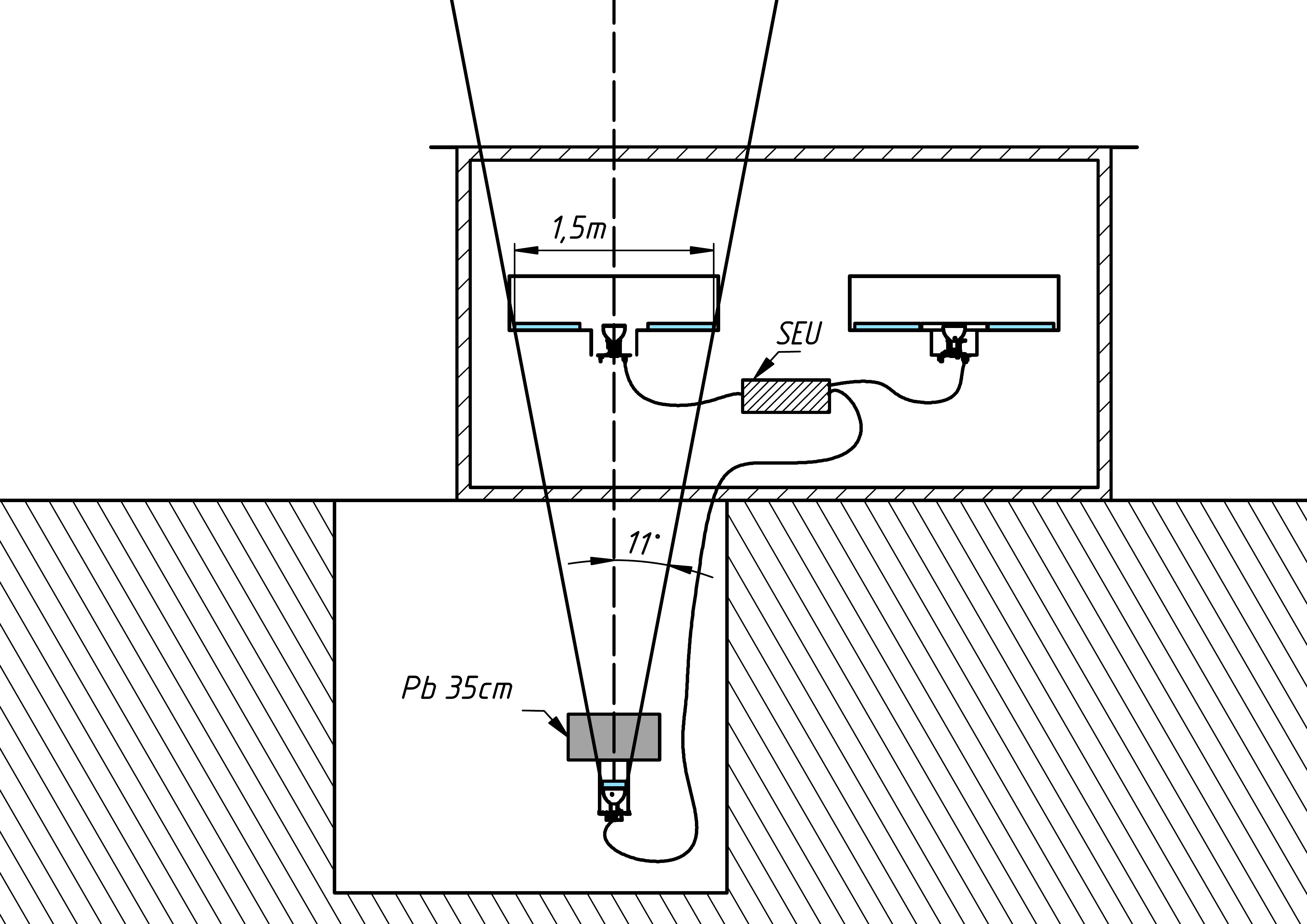}
    \caption{Measurement of the response spectrum in scintillation detector from vertical relativistic muons.}
    \label{fig:4}
\end{figure}

\noindent
Subsequently, in a shower arriving at zenith angle $\theta$ the density of particle flux through a detector of area $s$ at axis distance $r$ equals

\begin{align}
    \rhoY &= \frac{n(r,\theta)}{s(\theta)} =
    \frac{n(r,0\degr) \cdot \cos\theta}{s \cdot \cos\theta} = \nonumber \\
          &\quad = \frac{U(r)}{U_1(0\degr) / s} = 
          \frac{n(r)}{s}~\text{[m}^{-2}\text{].}
          \label{eq:4}
\end{align}

\noindent
It does not depend on the shower arrival angle since the amplitude of signal on the PMT anode does not change.

\subsection{Absolute Calibration of Detectors} \label{sec:2.4}

\begin{figure}[!htb]
    \centering
    \includegraphics[width=0.49\textwidth]{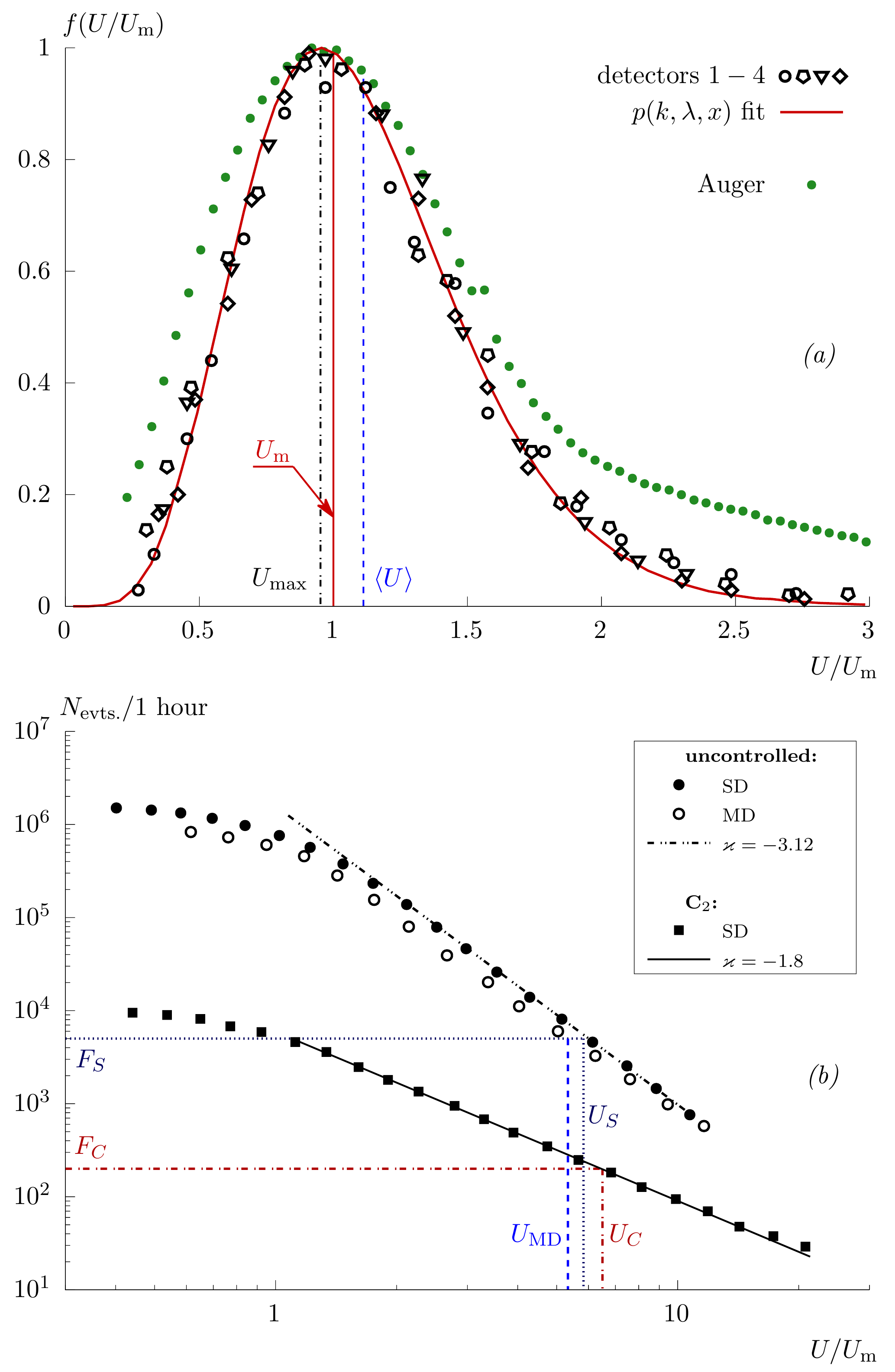}
    \caption{({\sl a}) Differential response distribution of a 2-\sqrm{} scintillation detector from vertical relativistic muons. Empty symbols denote different individual detectors, solid curve~--- approximation of these data with gamma-distribution \eqn{eq:5}. Calibration spectrum of the Auger detectors is shown with dark circles (data were extracted from Fig.~2 in work~\cite{bib:1}). ({\sl b}) Integral response spectra in scintillation detectors accumulated from cosmic muon background in different conditions: from uncontrolled surface detector, \CCTr{} spectrum and spectrum from uncontrolled detector inside a muon registration point. All spectra are normalized to 1~hour exposure.}
    \label{fig:5}
\end{figure}

The response distribution from a single vertical relativistic muon in a standard detector was obtained in a special experiment~\cite{bib:7}, its schematics is given in \fign{fig:4}. The experimental setup consisted of a standard surface-based detector station with two 2-\sqrm{} scintillation detectors operating in coincidence mode. The station was placed on top of a small 0.04~\sqrm{} controlling detector which was vertically aligned with one of the station detectors, the controlled one. This alignment formed a telescope selecting background particles with maximum deviation from vertical $\nthick\approx 11\degr$ which triggered both controlled and controlling detectors. The 35-cm layer of lead above the controlling detector absorbed soft particles and ensured triggering only from relativistic muons. The average coincidence rate was 0.5~s$^{-1}$. The resulted distributions of response from vertical muon for several detectors are shown in \fign{fig:5}{\sl a} with different empty symbols. Solid line represents approximation of these amplitude spectra with gamma-distribution:

\begin{equation}
    p(k, \lambda, x) = \lambda^k \cdot k^{-0.5} \cdot x^{k - 1} \cdot e^{-\lambda x}\text{.}
    \label{eq:5}
\end{equation}

\noindent
The dash--dotted vertical line indicates the maximum of this distribution ($U_{\text{max}} = (k - 1) / \lambda = 0.822$), solid vertical line~--- median ($U_{\text{m}} = 0.863$) and dashed vertical line~--- average amplitude ($\left<U\right> = k / \lambda = 0.96$). As a response unit at \yeas{} the median amplitude value $U_{\text{m}}$ was adopted. This choice was justified by the fact that during the initial years of \yeas{} operation, at the time when the experiment on the absolute calibration was conducted, the median of the spectrum was measured more precisely than the maximum.

Dark circles in \fign{fig:5}{\sl a} represent the calibration spectrum of Auger muon detector (see Fig.~2 in~\cite{bib:1}). The spectrum accumulates signals from 64 separate PMTs in a scintillator block, which possibly explains its wider distribution. The calibration level of Auger detectors corresponds to the maximum of distribution and within 5\% agrees with calibration level of \yeas{}.

\subsection{Operation Control and Calibration of Detectors} \label{sec:2.5}

It is impractical and virtually impossible to measure the vertical muon spectrum with a telescope for daily calibration and control of every scintillation detector of the array. But the background muon spectrum is well-suited for this purpose~\cite{bib:7}. The station electronics unit (SEU, see \fign{fig:4}) has a local trigger selecting the so-called ``double coincidences''~--- events from two detectors occurring within a 2-\usec{} interval (the \CCTr{} trigger). The SEU selects only \CCTr{} events, the amplitudes of which are digitized and stored in memory. These data are transferred from detector stations into the central registration system of the array which accumulates \CCTr{} spectra from every station. The frequency of a \CCTr{} spectrum amounts to $\nthick\sim 2-3$ events per second. The amplitude distribution from each detector in this case reflects the so-called density spectrum of low energy EASs. In integrated form this spectrum can be described with a power function $U^{-\varkappa}$ with index value $\varkappa \approx 1.5$ in a wide density range.

At low densities in a \CCTr{} spectrum there is an additional component arising from single low-energy particles falling within the 2-\usec{} interval due to random coincidences. Since time resolution of \CCTr{} trigger is $\tau = 2$~\usec{} and count rate is $N \approx 400$~s$^{-1}$, the number of random events in the spectrum of a typical station is rather high: $2 \times \tau \times N_1 \times N_2 \approx 0.6-0.7$~s$^{-1}$. They increase the steepness of a real distribution in operating range of densities from approximately 2 to 20 particles per detector. Experimentally measured steepness of the integrated spectrum within this range amounts to $\nmed\varkappa = 1.7 - 1.8$. The integral \CCTr{} spectrum is shown in \fign{fig:5}{\sl b}.

To make use of this spectrum for regular calibration of operating detectors, a different experiment was conducted (see \fign{fig:4}). Two spectra were simultaneously measured: spectrum of vertical muons via telescope, and \CCTr{} spectrum from two 2-\sqrm{} detectors in a station collected via SEU. For \CCTr{} spectrum a fixed frequency was chosen, $F_C = 200$~events per hour, and from the vertical muon spectrum a relation was established between the corresponding amplitude $U_C$ (calibration level) and single particle level $U_{\text{m}}$. This relation was measured for several scintillation detectors and for all of them the $\lg(U_C/U_{\text{m}}) = 0.81$ relation was established with a good agreement.

The muon registration points (and secondary non-triggering surface stations in the central region of the array) lack \CCTr{} selection circuits, all their detectors operate independently. Thus these detectors are calibrated by the uncontrolled integral spectrum with significantly higher frequency. In the amplitude range from $\nmed 1-2$ up to 10 particles this spectrum can be approximated with the same power law but with much steeper index, $\varkappa \approx 3.1$. The increased steepness mainly arises from the zenith-angular distribution of single muons. As a base frequency for such spectrum the value $F_S = 5000$~events per hour was adopted (see \fign{fig:5}{\sl b}). The relations between amplitudes that correspond to $F_S$ frequency ($U_S$ and $U_{\text{MD}}$ correspondingly) and the level of a single vertical muon $U_{\text{m}}$ were also determined in the experiment with telescope. The corresponding values are $\lg(U_S / U_{\text{m}}) = 0.77$ for surface detectors and $\lg(U_{\text{MD}} / U_{\text{m}}) = 0.725$ for muon detectors.

All spectra shown in \fign{fig:5}{\sl b} are normalized to 1~hour exposure, although the real exposure of \CCTr{} spectrum is 2 hours and both uncontrolled spectra are collected for 30~minutes. This provides the calibration accuracy $10-15$\%. All detectors are calibrated in a continuous cycle.

%% file: 03-auger.tex
\section{The Auger experiment} %\label{sec:3}

\subsection{Measurement of Muon Density} \label{sec:3.1}

The technique of muon density measurements in work~\cite{bib:1} is described
in Section~3.1\footnote{For PDF version see page 8, last paragraph in the left column.}, quote,

\begin{quotation}
    ``The expected number of muons, $\mu(r,E,\theta)$%
    % footnote
\footnote{Here and below in the quoted paragraph the correspondence between designations in this work and the original paper~\cite{bib:1}: number of muons $n$~--- $\mu$; primary energy $\E$~--- $E$; detector area $s$~--- $S$.},
    that hit a scintillation module located at a distance $r$ from the impact point of a shower impinging with zenith angle $\theta$ is then derived as $\mu(r,E,\theta) = \rho(r,E,\theta) S \cos\theta$, with $S \cos\theta$ the projected aperture of the detectors\ldots'',
\end{quotation}

\noindent
end quote. Let's write down the expression for estimating the expected number of registered muons from the quoted paragraph separately:

\begin{equation}
    n(r, \Esd, \theta) = \rho(r,\Esd,\theta) \cdot s \cdot\cos\theta\text{.}
    \label{eq:6}
\end{equation}

\begin{figure}[!htb]
    \centering
    \includegraphics[width=0.49\textwidth]{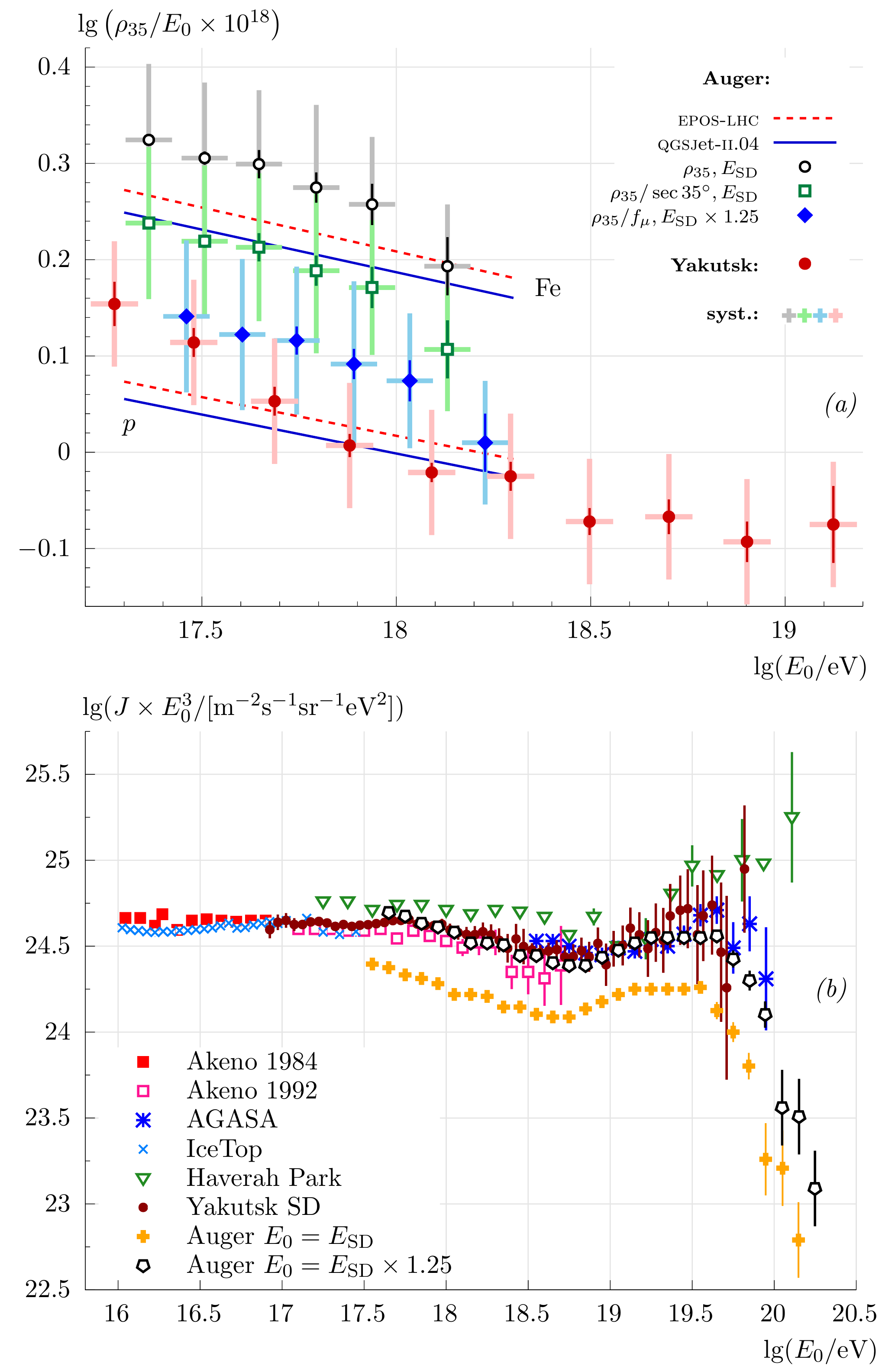}
    \caption{({\sl a}) Energy dependencies of the $\rhoPAO$ parameter obtained by the Auger Collaboration (see \fign{fig:1}) and \yeas{}. The ``$\rho_{35}/\sec35\degr,\Esd$'' data set represents muon densities from \fign{fig:1} reduced by factor $\sec{35\degr} = 1.22$, the ``$\rho_{35}/\fmu, \Esd \times 1.25$'' set~--- same densities reduced by factor $f_{\mu} = 1.22 \times 1.25 = 1.525$ with primary energy $\Esd$ increased by factor 1.25. ({\sl b}) Differential energy spectrum of PCR according to the data of various world arrays. The ``$\Esd \times 1.25$'' data set is the Auger spectrum~\cite{auger:icrc17} with primary energy increased by factor 1.25.}
    \label{fig:6}
\end{figure}

\noindent
Here $\Esd$ is the Auger CR energy estimation ($\E$) derived from readings of surface detectors (SD). From relation \eqn{eq:6} it follows that muon densities in work~\cite{bib:1} were determined according to the formula:

\begin{align}
    \rho(r,\Esd,\theta) &=
    \frac{n(r,\Esd,\theta)}{s \cdot \cos\theta} =
    \frac{n(r,\Esd,\theta)}{s} \cdot \sec\theta = \nonumber \\
    &\quad = \rhoY(r) \cdot \sec\theta~\text{[m],}
    \label{eq:7}
\end{align}

\noindent
where $\rhoY$ is the muon density defined by Eq.~\eqn{eq:4}, which is independent from shower zenith angle (see Section~2.3). Hence, it follows that $\rhoPAO$ values shown in \fign{fig:1} were overestimated by factor $\sec 35\degr \approx 1.22$. These values, reduced by 1.22 times, are shown in \fign{fig:6}{\sl a} with ``$\rhoPAO/\sec35\degr, \Esd$'' data set. Lines in this figure represent the results of calculations performed for Auger muon detectors with the use of \qgsii{}~\cite{QGSJetII} and \eposlhc{}~\cite{EPOSLHC} models for primary protons ($p$) and iron nuclei (Fe)~\cite{bib:1}.

Now the resulting muon densities could be more reasonably connected to heavy composition of primary CR (PCR). If this is the case, then the heavily discussed ``muon puzzle''~--- abnormally big discrepancy between theory and experiment~\cite{bib:2, WHISP:2021}~--- loses all its urgency. The only remaining disagreement here is conditioned, in our opinion, by primary energy estimation. Earlier, the PCR energy spectra measured by \yeas{} and the Auger Collaboration were compared~\cite{bib:9}. They drastically differ from each other. We have already shown that this disagreement could be mitigated by increasing the $\Esd$ energy estimation by factor 1.25~\cite{bib:10}. In \fign{fig:6}{\sl b} the differential PCR energy spectrum is shown according to several world experiments: Akeno (1984, 1992)~\cite{bib:11, bib:12}, AGASA~\cite{bib:13}, IceTop~\cite{bib:14}, Haverah Park array (HP)~\cite{bib:16} and Auger~\cite{auger:icrc17}. The \yeas{} spectrum was obtained from SD readings~\cite{bib:9}. The ``$\Esd \times 1.25$'' data set represents the Auger spectrum~\cite{auger:icrc17} with primary energy increased by factor 1.25.

As a result of energy rescaling of the Auger data, all densities presented in \fign{fig:6}{\sl a} should go down by 25\% (presented by the ``$\rhoPAO / \fmu, \Esd \times 1.25$'' data set). The two unidirectional factors lead to the total decrease of the measured densities by factor $\fmu$:

\begin{equation}
    f_{\mu} = \left<\rho\right>_{\text{exp}} / \left<\rho\right>_{\text{sim}}
    = 1.22 \times 1.25 = 1.525\text{,}
    \label{eq:8}
\end{equation}

\noindent
where ``exp'' and ``sim'' indicate, accordingly, experimental data and results of simulation performed with the use of the \qgsii{} model~\cite{bib:1}. Our estimation agrees with the value $\fmu = 1.53 \pm 0.13$(stat) obtained by authors of the work~\cite{bib:1}.

\subsection{Primary Energy Estimation} \label{sec:3.2}

\begin{figure}[!htb]
    \centering
    \includegraphics[width=0.49\textwidth]{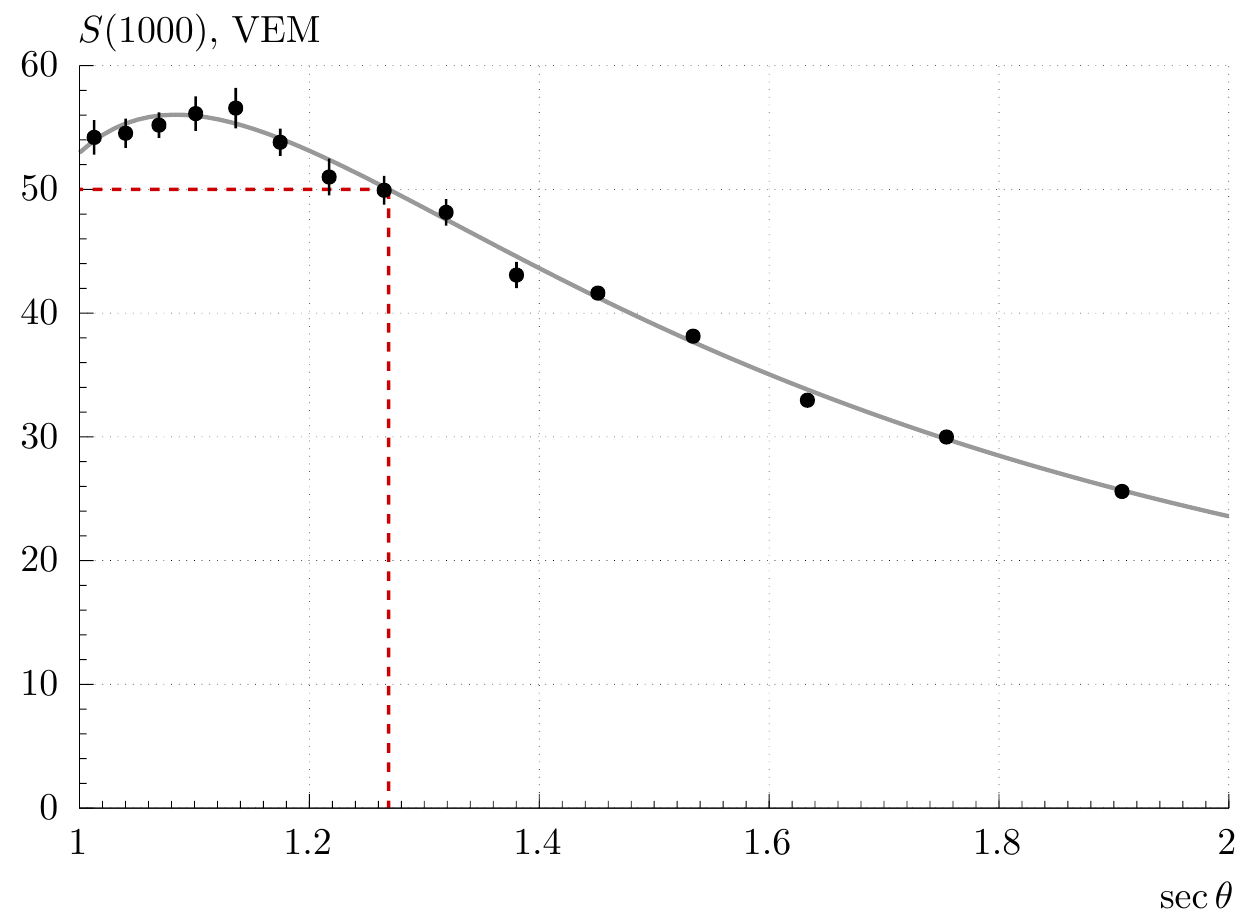}
    \caption{The attenuation curve $\fattb$ \eqn{eq:10} of the Auger SD signal at 1000~m from shower axis $S(1000)$. The reference angle $\theta_{\text{med}} = 38\degr$ is shown with vertical dashed line, the corresponding signal $S_{38} \approx 50$~VEM~--- with horizontal dashed line. Data were extracted from Fig.~40 in work~\cite{bib:17}.}
    \label{fig:7}
\end{figure}

The fundamental formula for primary energy estimation in the Auger experiment is given in~\cite{bib:17}:

\begin{equation}
    \Esd = A (S(1000) / \fattb)^B~\text{[eV],}
    \label{eq:9}
\end{equation}

\noindent
where $A = (1.90 \pm 0.05) \times 10^{17}$~eV and $B = 1.025 \pm 0.007$. The energy $\Esd$ is determined from the shower classification parameter $S(1000)$~--- particle density measured with SD at 1000~m from shower axis and recalculated to the median zenith angle $\theta_{\text{med}} = 38\degr$, $S_{38} = S(1000)/\fattb$ with the use of the attenuation curve shown in \fign{fig:7}:

\begin{equation}
    \fattb = 1 + ax + bx^2 + cx^3\text{,}
    \label{eq:10}
\end{equation}

\noindent
where $x = \cos^2\theta - \cos^2 38\degr$; $a = 0.980 \pm 0.004$, $b = -1.68 \pm 0.01$ and $c = -1.30 \pm 0.45$. Here $S_{38}$ is the energy estimator.

The Auger SD is a water tank-based Cherenkov detector (WCD) with radius 3.6~m, area $s = 10.2$~\sqrm{} and height 120~cm. The $S(1000)$ parameter equals the total response from charged EAS particles in SD measured in units of vertical equivalent muon (VEM). 1~VEM equals the energy deposit of a single vertical relativistic muon traversing 120~cm of water in a tank. The energy is spent on ionization: 120~\depth{} $\times$ 2.22~MeV cm$^2$/g = 266~MeV. Charged particles with energy below 1~VEM never reach the bottom of WCD. The detector measures not ionization losses per se, but only a fraction emitted in a form of Cherenkov light generated by relativistic charged particles in water. The resulting photon gas reflects from inner surfaces of the tank and is registered by three PMTs mounted at the top. The most important fact here is that the number of generated photons critically depends on the energy of charged particles.

Prototypes of Auger SDs were WCDs used at HP array. Some calculation results of the energy deposit in HP tanks are listed in Table~\ref{t:1}~\cite{bib:18}. Threshold rapidity of Cherenkov emission in water is $\beta = 0.75$, which corresponds to kinetic energy 150~MeV for muons and 260~keV for electrons. It follows that a muon with energy equivalent of 1~VEM generates Cherenkov light only in the initial section of the track, until its energy degrades below 150~MeV. For a full-fledged light track the muon energy must exceed 416~MeV (150 + 266~MeV). Measurements performed at HP have demonstrated that when the energy of a single vertical muon exceeded 500~MeV, the amplitude of signal at the PMT in WCD stopped changing~\cite{bib:18}.

\input{table1}

\subsection{Calculation of Energy Deposit in Detectors} \label{sec:3.3}

As was mentioned above, WCDs used at HP were prototypes of Auger SDs. The PCR energy spectrum obtained at HP~\cite{bib:16} is shown in \fign{fig:6}{\sl b}: it is higher than both \yeas{} and AGASA~\cite{bib:13} spectra, but is significantly higher than the Auger spectrum~\cite{auger:icrc17}. The source of such a discrepancy is puzzling, given that detectors of HP and Auger share similar design: both use a 120-cm layer of water as a light radiator and it is hard to admit that intensities of the resulting spectra would differ so drastically. The reason is, most likely, in different estimations of showers energy. To investigate this we have performed a series of simulations with \corsika{} code~\cite{CORSIKA} using the \qgsii{} model for primary protons with energy $10^{19}$~eV for three experiments: \yeas, HP and Auger. In each case the responses of \yeas{} scintillation detectors were calculated according to the technique described in~\cite{bib:9, bib:21, bib:22}.

\begin{figure}[!htb]
    \centering
    \includegraphics[width=0.49\textwidth]{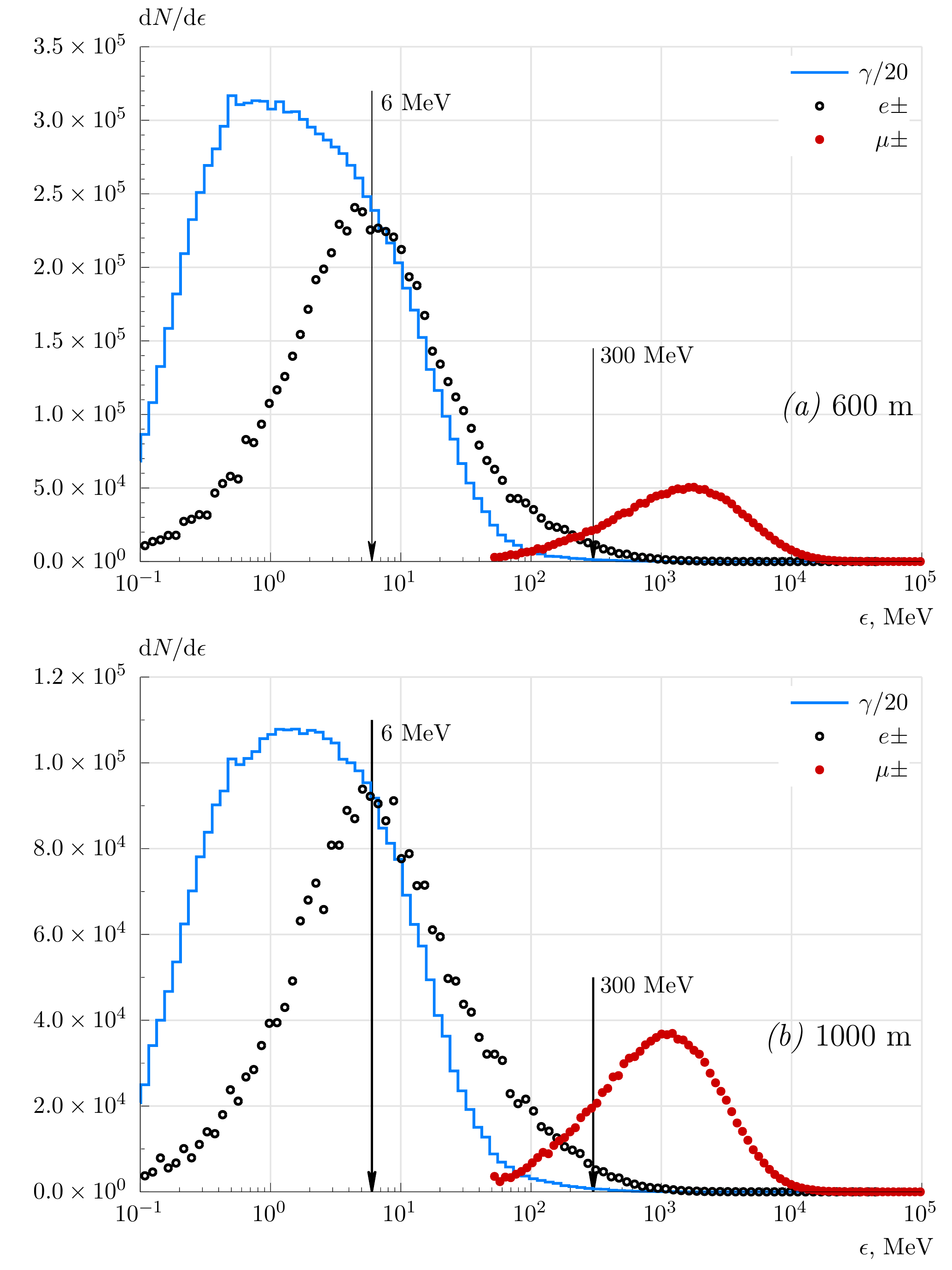}
    \caption{Energy spectra of EAS particles at $r = 600$~m ({\sl a}) and $r = 1000$~m ({\sl b}) from the axis calculated within the framework of \qgsii{} model for primary protons with $\E = 10^{19}$~eV and $\cos\theta=0.8$ for conditions of the Auger experiment (the plotted intensity of photon flux is downscaled by factor 20). In both cases a detector is a ring around shower axis in the plane of shower front, with inner and outer radii $\lg(r / \text{m}) \pm 0.02$, which is projected to the observation plane (see~\cite{bib:5, bib:6, bib:10} for details). The values 6~MeV and 300~MeV represent, correspondingly, the real cover thickness of the \yeas{} SD and the nominal threshold of Auger.}
    \label{fig:8}
\end{figure}

Energy spectra of EAS particles calculated for conditions of Auger are shown in \fign{fig:8} at $r = 600$~m ({\sl a}) and $r = 1000$~m ({\sl b}) from shower axis (the displayed number of photons is downscaled by 20 times). The 300~MeV threshold doesn't mean that the Auger SD registers only particles with energy above this value. All particles passed through the cover of a detector (we assume its thickness is 9~\depth) lose energy on ionization, but not all of them produce a Cherenkov track. In fact the Auger SD acts selectively to the kind of particles, especially to high-energy photons in electromagnetic EAS component. Due to this feature of the Auger SD it is hard to understand how actually the response $S_{38} \equiv S(1000) / \fattb$ is formed. This substantially complicates the interpretation of experimental data and the detector simulation. In this sense the \yeas{} SD is much simpler: all particles with energy above 17.75~MeV (6 + 11.75~MeV) manifest themselves as full-fledged responses, proportional to 1~VEM = 11.75~MeV.

One of the aims of the simulation was to check the coupling coefficients $A$ in expression for primary energy estimation by the energy estimator $\RhoC$ at different arrays:

\begin{equation}
    \E = A \times \RhoC^B~\text{[eV],}
    \label{eq:11}
\end{equation}

\noindent
where $\RhoC$ is the density of nominal particles measured at axis distance $\Rc$ in units of VEM per 1~\sqrm{} of the detector area. The results are listed in Table~\ref{t:2}. In column~2 the atmospheric depth above the given array $\xobs$ is listed, in column~3~--- axis distance $\Rc$ where the classification parameter is measured, in column~4~--- threshold of particle cut-off $\Ecut$ (see \fign{fig:8}); particles with energies below $\Ecut$ were discarded from further treatment. Column~6 lists the zenith angle $\ZenC$ to which all individual densities were converted, column~7~--- values of the coefficient $A$ in relation \eqn{eq:11}. The power parameter $B$ was assumed equal to $B = 1.025 \pm 0.003$.

Parameters of relation \eqn{eq:11} determined from the Yakutsk experimental data~\cite{bib:9} are listed in row~1 of the table; values obtained in this work from simulation results~--- in row~2. These values agree with each other within 3\%, indicating that simulation and experiment were adequate to each other.

With this toy model one can run a ``diagnostic test'' for any other array of interest. In row~4 experimental data of HP are shown~\cite{bib:18, bib:19}. This array was located at the depth $\xobs = 1016$~\depth, close to the \yeas{} level. In both experiments the same EAS energy estimator was used~--- particle density $\rhosSOOV$ converted to $\theta = 0\degr$ with the formula:

\begin{gather}
    \rhosSOOV = \rhosSOOT \times \nonumber \\
    \times \exp \left[
        \frac{(\sec\theta - 1) \cdot \xobs}{\lambda}
    \right]~\text{[m}^{-2}~\text{],}
    \label{eq:12}
\end{gather}

\noindent
with attenuation lengths $\lambda = 500 \pm 30$~\depth~\cite{bib:9} and $\lambda = 760 \pm 40$~\depth~\cite{bib:19} correspondingly. If we apply relation \eqn{eq:12}, then according to \eqn{eq:11} we obtain the value listed in row~5 ($A / 10^{17}~\text{eV} = 4.91 \pm 0.02$). It is $\nthick\approx 1.43$ times lower than experimental value~\cite{bib:18, bib:19} in row~4 ($A / 10^{17}~\text{eV} = 7.04$) and disagrees with PCR energy spectrum displayed on \fign{fig:6}{\sl b}. In row~6 another result for HP is shown, where the measured densities $\rhosSOOT$ were recalculated to zenith angle $\theta = 38\degr$ instead of vertical direction. In this case the resulting coefficient ($A / 10^{17}~\text{eV} = 6.91 \pm 0.02$) is consistent with the value obtained in experiment~\cite{bib:18, bib:19} ($A / 10^{17}~\text{eV} = 7.04$) in row~4.

It is not entirely clear how the zenith angle value $\theta = 38\degr$ appeared here; there was no mention about it in the HP papers. Thus, we had no other option but to turn to simulations. In row~3 a variant of \yeas{} energy calibration is shown according to relations \eqn{eq:11} and \eqn{eq:12}, but with particle densities converted to $\ZenC = 38\degr$. In such a case the energy estimators of two arrays differ by factor $\nthick\approx 1.26$ ($17.4~\text{m}^{-2} / 13.8~\text{m}^{-2}$). This is due to the fact that, as was described above, 26\% of shower particles do not manifest themselves in any way when traversing through a WCD. The value $\sec 38\degr \approx 1.27$ deserves special attention~--- is it just a random coincidence? Actually no: row~7 shows a case when ``the number of complete responses'' is calculated from all particles with energies above 300~MeV. Together with \fign{fig:8} these data demonstrate that if a shower arrives at zenith angle $\theta \approx 38\degr$ then in the range between 6~MeV (row~3) to 300~MeV (row~7) there are nearly 58\% $\left(\frac{17.4~\text{m}^{-2} - 11.0~\text{m}^{-2}}{11.0~\text{m}^{-2}} \times 100\%\right)$ of ``incomplete'' responses with values ranging from 0 to 1~VEM. It is these events that lead to incorrect estimation of PCR energy according to Eq. \eqn{eq:9}.

\input{table2}

Now let's take a look at the Auger experiment~\cite{auger:icrc17}. If we multiply the energy estimator $\rho(1000,38\degr) = 4.90 \pm 0.01$~m$^{-2}$ in row~8 by the SD area 10.2~\sqrm{}, then we obtain the total response $S_{38} = 49.98 \pm 0.10 \approx 50$~VEM which is a reference signal shown in \fign{fig:7}. Row~9 lists the results of calculation for the \yeas{} SD as if it was placed side by side with the Auger SD. The resulting values nearly match with the results presented in row~8. But the signal of the \yeas{} SD is 26\% higher than signals of both HP and Auger detectors. Hence, Eq. \eqn{eq:9} is not applicable to these arrays, as it doesn't account for 26\% of all particles that are ``invisible'' to a WCD. Row~10 lists the result obtained with the same values of $\Ecut$ and $\ZenC$ as in row~6 for HP. If one adopts this calibration result for calculation of the total response $S(1000)$ according to Eq. \eqn{eq:9}, then the value of coupling coefficient should be:

\begin{equation}
    A = (2.38 \pm 0.05) \times 10^{17}~\text{[eV].}
    \label{eq:13}
\end{equation}

\noindent
The correctness of this hypothesis is supported by simulation results performed by the Auger Collaboration listed in row~11\textsuperscript{*}~\cite{bib:20}. The resulting energy estimator, $\rho(1000,38\degr) = 3.82 \pm 0.02$~m$^{-2}$, differs from the corresponding value in row~10, $\rho(1000,38\degr) = 3.92 \pm 0.02$~m$^{-2}$, by 2.5\%. This simulation was performed within the framework of the \qgsii{} model for primary protons with energy $10^{19}$~eV. The total response $S_{38} \equiv 39$~VEM in this work, according to formula \eqn{eq:9}, corresponds to the energy $\Esd \approx 8.1 \times 10^{18}$~eV, which is $\nthick\approx 1.24$ times lower than the value initially set in the simulation~\cite{bib:20}. To obtain the value $\Esd = 10^{19}$~eV the coupling coefficient $A \approx (1.90 \times 1.24) \times 10^{17} \approx 2.36 \times 10^{17}$~eV should be used, which is close to the value given in \eqn{eq:13}.

%% file: table1.tex
\begin{table}[!htb]
    \centering
    \caption{Energy losses of vertical muons and full number of emitted Cherenkov photons in 120~cm of water.}
    \label{t:1}
    \setlength{\tabcolsep}{0.7em}     % wider columns padding
    \renewcommand{\arraystretch}{1.2} % wider rows padding
    \newcommand{\COne}{0.07\textwidth}
    \newcommand{\CTwo}{0.13\textwidth}
    \newcommand{\CThree}{0.10\textwidth}
    \newcommand{\CFour}{0.06\textwidth}
    \begin{tabular}{%
        R{\COne}%
        C{\CTwo}%
        R{\CThree}%
        R{\CFour}%
    }
        \hline
        \hline
        \multicolumn{1}{C{\COne}}{Kinetic energy $T$, MeV} &
        Energy deposit in 120~cm of water $\Delta T$, MeV &
        \multicolumn{1}{C{\CThree}}{Full number of emitted photons, $N_{\gamma}$} &
        \multicolumn{1}{C{\CFour}}{$N_{\gamma} / \Delta T$, MeV$^{-1}$} \\
        \hline
            100 & 100 &  1480 &  15 \\
            200 & 200 &  9200 &  46 \\
            300 & 220 & 19000 &  86 \\
            400 & 220 & 22000 &  97 \\
            500 & 220 & 23000 & 110 \\
        $\nthick>1000$ & 220 & 25000 & 114 \\
        \hline
        \hline
    \end{tabular}
\end{table}

%% file: table2.tex
\begin{table*}[htb]
    \centering
    \caption{Parameters of the relation for PCR energy estimation via classification parameter $\RhoC$ \eqn{eq:11} obtained for \yeas{}, HP and Auger (column~8 indicates how the resulting values were obtained: either from published experimental data or simulations performed for the \yeas{} SD placed in conditions of a given experiment; except for row~11\textsuperscript{*}, which lists the result given in~\cite{bib:20}; rows and columns were numbered for convenience)}
    \label{t:2}
    \setlength{\tabcolsep}{0.7em}     % wider columns padding
    \renewcommand{\arraystretch}{1.2} % wider rows padding
    \newcommand{\CTwo}{0.04\textwidth}
    \newcommand{\CThree}{0.04\textwidth}
    \newcommand{\CFour}{0.04\textwidth}
    \newcommand{\CSix}{0.08\textwidth}
    \newcommand{\CSeven}{0.09\textwidth}
    \newcommand{\CEight}{0.12\textwidth}
    \begin{tabular}{%
        r%
        R{\CTwo}%
        R{\CThree}%
        R{\CFour}%
        r%
        R{\CSix}%
        R{\CSeven}%
        L{\CEight}%
        l%
    }
        \hline
        \hline
        1                     & \multicolumn{1}{c}{2} & \multicolumn{1}{c}{3} &
        \multicolumn{1}{c}{4} & \multicolumn{1}{c}{5} & \multicolumn{1}{c}{6} &
        \multicolumn{1}{c}{7} & \multicolumn{1}{c}{8} & \multicolumn{1}{c}{9} \\
        \hline
                              &
        \multicolumn{1}{C{\CTwo}}{$\xobs$, \depth} &
        \multicolumn{1}{C{\CThree}}{$\Rc$, m} &
        \multicolumn{1}{C{\CFour}}{$\Ecut$, MeV} &
        \multicolumn{1}{c}{$\ZenC$} &
        \multicolumn{1}{C{\CSix}}{$\RhoC$, m$^{-2}$} &
        \multicolumn{1}{C{\CSeven}}{$A$ $\times 10^{17}$~eV} &
        \multicolumn{1}{C{\CEight}}{Source, $10^{19}$~eV} &
        Array [reference]\\
        \hline
         1 & 1020 & 600 & $6$ & $0\degr$ & $24.6 \pm 0.03$ &
            $3.76 \pm 0.30$ & experiment & \yeas~\cite{bib:9} \\
         2 & 1020 & 600 & $6$ & $0\degr$ & $25.4 \pm 0.02$ &
            $3.64 \pm 0.02$ & \qgsii, $p$ & \yeas{} \\
         3 & 1020 & 600 & $6$ & $38\degr$ & $17.4 \pm 0.02$ &
            $5.34 \pm 0.02$ & \qgsii, $p$ & \yeas{} \\
         4 & 1016 & 600 & $9$ & $0\degr$ & $13.6 \pm 0.02$ &
            \multicolumn{1}{c}{7.04} & experiment & HP~\cite{bib:18, bib:19} \\
         5 & 1016 & 600 & $9$ & $0\degr$ & $19.0 \pm 0.02$ &
            $4.91 \pm 0.02$ & \qgsii, $p$ & HP \\
         6 & 1016 & 600 & $9$ & $38\degr$ & $13.8 \pm 0.02$ &
            $6.91 \pm 0.02$ & \qgsii, $p$ & HP \\
         7 & 1016 & 600 & $300$ & $38\degr$ & $11.0 \pm 0.02$ &
            $8.56 \pm 0.02$ & experiment & HP \\
         8 & 875 & 1000 & $9$ & $38\degr$ & $4.90 \pm 0.01$ &
            $19.0 \pm 0.05$ & experiment & Auger~\cite{auger:icrc17} \\
         9 & 875 & 1000 & $6$ & $38\degr$ & $5.02 \pm 0.02$ &
            $19.13 \pm 0.02$ & \qgsii, $p$ & Auger \\
        10 & 875 & 1000 & $9$ & $38\degr$ & $3.92 \pm 0.02$ &
            $23.75 \pm 0.02$ & \qgsii, $p$ & Auger \\
        \hline
        11\textsuperscript{*} & 875 & 1000 & $3$ & $38\degr$ &
            $3.82 \pm 0.02$ & $25.30 \pm 0.02$ & \qgsii, $p$ &
            Auger~\cite{bib:20} \\
        \hline
        \hline
    \end{tabular}
\end{table*}

%% file: 04-finale.tex
\section{Conclusion}

The combined analysis of methods for measuring the muon density at different EAS arrays made it possible to solve the problem of muon excess in the data of the Auger experiment~\cite{bib:2}. In Table~\ref{t:3} the values of $\fmu$ coefficients \eqn{eq:9} are shown, which are relations between muon densities in EAS obtained from experimental data and the values predicted by \qgsii{} model for primary protons~\cite{bib:1, bib:20}. Our estimations of these coefficients together with their constituents are also given. The agreement between these data sets suggests that the nature of the ``Muon puzzle'' is a methodical one. It is conditioned purely by fundamental physical limitation of WCDs used at the Auger array (and also of their prototypes at HP), namely~--- the ability to measure only a 3/4 of total ionization losses of EAS particles. On the one hand, this led to incorrect formula \eqn{eq:6} which overestimates the measured muon densities by factor $\sec35\degr$ in~\cite{bib:1} and by factor $\sec38\degr$ in~\cite{bib:20}. On the other hand, the 25\% under-measuring of total ionization losses (not accompanied by Cherenkov light emission) by Auger SDs resulted in underestimation of primary cosmic ray energy with formula \eqn{eq:9} by factor $\sec38\degr$. In this case formula \eqn{eq:9} should be used with the value of coupling coefficient $A$ given by \eqn{eq:13}. We will continue our investigation of this problem.

\input{table3}

%% file: table3.tex
\begin{table}[!htb]
    \centering
    \caption{The relation $\fmu = \left<\rho\right>_{\rm exp} / \left<\rho\right>_{\rm sim}$ between muon densities obtained from experimental data (exp)~\cite{bib:1, bib:20} and simulation results (sim) (the ``this work'' reference denotes our calculations for conditions of the Auger array)}
    \label{t:3}
    \setlength{\tabcolsep}{0.7em}     % wider columns padding
    \renewcommand{\arraystretch}{1.2} % wider rows padding
    \begin{tabular}{clc}
        \hline
        \hline
        Energy & Reference & $\fmu$ \\
        \hline
        $10^{18.0}$~eV & \cite{bib:1}  & $1.53 \pm 0.13$(stat) \\
          & this work & $1.549 = \sec35\degr \times \sec38\degr$ \\
        \hline
        $10^{19.0}$~eV & \cite{bib:20} & $1.61 \pm 0.21$(stat) \\
          & this work & $1.613 = \sec38\degr \times \sec38\degr$ \\
        \hline
        \hline
    \end{tabular}
\end{table}